\documentclass[useamsfonts]{pasj00}
\draft
\usepackage{multirow}

\begin{document}
\SetRunningHead{H. Yoshitake et al.}{Suzaku Observation of the Lockman Hole}

\title{	Long Term Variability of  O\emissiontype{VII} Line Intensity toward the Lockman Hole\\ Observed with Suzaku from 2006 to 2011}

\author{
Hiroshi \textsc{Yoshitake}
\thanks{Present Address is Central Research Laboratory, Hitachi, Ltd., 1-280,
Higashi-Koigakubo, Kokubunji, Tokyo 185-8601}, 
Kazuhiro \textsc{Sakai}, 
Kazuhisa \textsc{Mitsuda}, 
Noriko Y. \textsc{Yamasaki}, 
\\ Yoh \textsc{Takei}, 
and Ryo \textsc{Yamamoto}
} 

\altaffiltext{}{Institute of Space and Astronautical Science, Japan Aerospace Exploration Agency (ISAS/JAXA), 
\\ 3-1-1 Yoshinodai, Chuo, Sagamihara, Kanagawa 229-8510}
\email{sakai@astro.isas.jaxa.jp}



\maketitle

\begin{abstract}
Long-term time variabilities of the O\emissiontype{VII} (0.57 keV) emission in
the soft X-ray diffuse background were studied using six Suzaku annual
observations of blank sky towards the Lockman Hole made from 2006 to 2011.
After time intervals in which the emission was enhanced on time scales of a few
tens of ks were removed, the O\emissiontype{VII} intensity was found to be
constant from 2006 to 2009 within the 90\% statistical errors. The intensity in
2010 and 2011 was higher by $2-3$ LU ($= {\rm photons\ s}^{-1}{\rm cm}^{-2}{\rm
sr}^{-1}$) than the earlier values. The most plausible origin of the fast
variable component is Solar wind charge exchange (SWCX). The intensity increase
is not positively correlated with the proton flux at the L1 point. Since all
the observations were made in the same season of a year, the variation cannot
be explained by parallax of the SWCX induced X-ray emission from the
Heliosphere. We consider that it is related to the geometrical change of slow
and fast solar wind structures associated with the 11 year solar activity. The
observed variation was compared with that expected from the SWCX induced X-ray
emission model.
\end{abstract}

\section{Introduction}
\label{sec:intro}

The charge exchange process of solar wind ions with neutrals (SWCX) is
considered to contribute to the diffuse X-ray background, especially at
energies under 1 keV (\cite{kout06}). When solar wind ions interact with
neutral atoms (mainly H and He), electrons bound in the neutral are transferred
to the excited state of the ion, and move to the ground state by emitting X-
rays corresponding to the de-excitation energy. Thus the SWCX consists of
emission lines, among which the O\emissiontype{VII} triplet at 0.57 keV is the
most prominent and has been observed with X-ray CCD spectrometers on board
Chandra, XMM-Newton and Suzaku (\cite{snowden04}, \cite{fujimoto07}).

The SWCX-induced diffuse emissions are considered to arise from collisions with
neutrals of two different origins. One is the Earth's geocorona, thus the
emission is called ``geocoronal SWCX''. According to \citet{estrg03}, the
geocorona extends with a scale height of $ \sim8.2$ $R_{\rm E}$ (earth radii).
The other source is atoms of the local interstellar medium (LISM) flowing into
the interplanetary space (heliosphere) due to the relative motion of the solar
system against the LISM. The heliosphere extends up to $\sim$100 AU from the
Sun \citep{lallement05}, and the emission is called ``heliospheric SWCX''.

The contribution of the geocoronal SWCX to the ROSAT All Sky Survey (RASS) 3/4
keV band map was pointed out by \citet{cox98}, as an enhancement characterized
by variability of timescales of a few tens ks to a few days. XMM discovered
strong line emissions correlated with the solar wind conditions around the
Earth (\cite{snowden04}). Careful analysis might be able to remove this
variable emission from the other X-ray diffuse emission (eg. \cite{yoshino09}).
The general contribution of the heliospheric SWCX has been also debated and
demonstrated in several cases (\cite{kout07}, \cite{kout11}). In particular,
\citet{kout07} showed that several observed variations on short or medium time
scale can be interpreted to be of the heliospheric SWCX origin.

If the heliospheric SWCX actually contributes to a certain amount of the soft
X-ray background, long term emission variations related to the 11-year solar
cycle and the inhomogeneity of the interstellar neutral distributions around
the Sun are expected. Large variations are mainly expected because of the
anisotropic slow/fast solar wind distributions in the solar corona
(\cite{mccomas08}). Due to the difference in plasma temperature, the O$^{+7}$
and O$^{+8}$ ionization fractions of the slow solar wind (higher temperature)
are much larger than those of the fast solar wind. Conversely, the ionization
fractions of oxygen ions with ionization state lower than O$^{+7}$ are higher
in the fast solar wind. During the solar minimum phase, the slow wind is
restricted to within a low heliocentric ecliptic latitude area
$|\beta|\lesssim 20^{\circ}$, and the fast wind widely extends from the polar
regions (\cite{mccomas08}). On the other hand, the slow wind covers almost the
whole solar corona during the maximum phase. For this reason, the heliospheric
SWCX-induced emission at high ecliptic latitudes is expected to depend on the
long term (11-year) solar activity.

\citet{kout06} calculated all sky emission maps of the heliospheric SWCX
induced O\emissiontype{VII} and O\emissiontype{VIII} lines on the basis of
averaged, minimum and maximum solar activity in Solar Cycle 22. They obtained
different O\emissiontype{VII} and O\emissiontype{VIII} maps with different solar
activity levels, especially toward the high ecliptic latitudes. According to
these model simulations, O\emissiontype{VII} at higher ecliptic latitudes may
be enhanced by 1--2 Line Unit (LU $= {\rm photons\ s}^{-1}{\rm cm}^{-2}{\rm
sr}^{-1}$) at solar maximum relative to minimum from the same observing site
along the Earth orbit.

\begin{table*}[ht]
\begin{center}
\caption{Log of the Suzaku Lockman Hole observations.}
\begin{tabular}{ccccccccc}
\hline \hline
ID  & Start / End date (UT) & \multicolumn{2}{c}{Exposure (ks)} &  \multicolumn{2}{c}{Screening} 
& \multicolumn{2}{c}{Pointing} & Roll angle \\ 
 & YYMMDD hh:mm:ss  & total$^\ast$ & screened$^\dagger$ & 1$^\ddagger$ & 2$^\S$ & R. A. & Dec. & \\
\hline
LH06 &  060517 17:44:06 /  060519 19:03:18 & 80.4 & 32.5 & $\checkmark$ & & 162$^{\circ}$.937 & 57$^{\circ}$.256 & 281$^{\circ}$.872 \\
LH07 &  070503 23:12:08 /  070506 02:00:19 & 96.1 & 56.7 & & $\checkmark$ & 162$^{\circ}$.937 & 57$^{\circ}$.258 & 319$^{\circ}$.511 \\
LH08 &  080518 11:07:29 /  080520 01:16:14 & 83.4 & 58.4 & & & 162$^{\circ}$.937 & 57$^{\circ}$.255 & 281$^{\circ}$.530 \\
LH09 &  090612 07:17:40 /  090614 01:31:21 & 92.8 & 63.8 & & & 162$^{\circ}$.938 & 57$^{\circ}$.255 & 281$^{\circ}$.530 \\
LH10 &  100611 07:29:06 /  100613 01:59:22 & 78.0 & 50.0 & $\checkmark$ & & 162$^{\circ}$.937 & 57$^{\circ}$.251 & 279$^{\circ}$.887 \\
LH11 &  110504 17:46:34 /  110505 18:25:20 & 42.3 & 19.4 & & $\checkmark$ & 162$^{\circ}$.920 & 57$^{\circ}$.251 & 305$^{\circ}$.984 \\
\hline
\multicolumn{9}{l}{\small $^\ast$ 
 Total exposure of the XIS1 after the standard screening. } \\ 
\multicolumn{9}{l}{\small $^\dagger$ 
 Screened exposure extracted when COR2 $>$ 8 GV $c^{-1}$ and screened by the criteria 1 and 2 (next 2 colums). } \\
\multicolumn{9}{l}{\small $^\ddagger$ 
  Effect of the geocoronal SWCX induced emission was removed (see $\S$\ref{subsec:gswcx}).}\\
\multicolumn{9}{l}{\small $^\S$ 
  Contamination of the neutral O\emissiontype{I} K$\alpha$ emission was removed (see $\S$\ref{subsec:neu_O}). }
\end{tabular}
\label{tab:obslog}
\end{center}
\end{table*}

The purpose of this paper is to investigate the variabilities of the
heliospheric SWCX induced O\emissiontype{VII} line at high ecliptic latitudes,
related to the long term solar cycle of $\sim11$ years. To avoid the direction
dependence of heliospheric SWCX, we must use a set of observations looking
toward the same direction. We also need to note that parallax also produces
variations, since the distributions of interstellar neutrals in the heliosphere
are quite anisotropic (\cite{pepino04}). We thus need to use those observations
made at the same period of time in each year. In addition, it is required to
remove time intervals during which the O\emissiontype{VII} intensity is
enhanced over $\sim$tens of ks, since this indicates enhancement by geocoronal
SWCX.
 
We analyzed O\emissiontype{VII} and O\emissiontype{VIII} line intensities
toward the Lockman Hole (in the high ecliptic latitude area at $\beta =
45.2^{\circ}$) with Suzaku (\cite{mitsuda07}), observed annually from 2006 to
2011. This interval covers about half a Solar Cycle headed from the solar
minimum phase at the end of Cycle 23 to the maximum of Cycle 24. This is the
first time that the energy spectra from a blank field region, with strictly
fixed line of sight and observation configuration, have been compared. If
significant differences of the oxygen line intensity still exist, these
differences will indicate the long term variability of the heliospheric SWCX
induced emission.

In this paper, the stated error ranges show 90\% confidence level from the
center value, and the vertical error bars in the figures indicate $\pm 1\sigma$
level unless otherwise noted.

\section{Observations}
\label{subsec:obs}

Suzaku has annually observed the Lockman Hole (hereafter LH) direction
$(\alpha, \delta) / (\ell, b)= (162^{\circ}.94,\,
57^{\circ}.25)/(149^{\circ}.71,\, 53^{\circ}.21)$ since 2006 as a calibration
target for the Hard X-ray Detector (HXD). This field is characterized by very
low hydrogen column density; $N_{\rm H} = 5.6\times10^{19}\ {\rm cm}^{-2}$
(\cite{dickey90}, and \cite{hasinger93}), or $5.8\times10^{19}\ {\rm cm}^{-2}$\
(\cite{kalb05}). Details of the observations are summarized in Table
\ref{tab:obslog}. From this point forth we will refer to the individual
observations by using the LH acronym and the corresponding year as listed in
the first column of Table \ref{tab:obslog}. Although all observations faced in
the same pointing direction within $2'$, the roll angles were different by up
to $\sim 40^{\circ}$ (the fields of view differ by $\sim 20\%$). We use the
data of the X-ray Imaging Spectrometer (XIS) on--board Suzaku (\cite{koyama07})
ver.2.0, 2.0, 2.2, 2.4, 2.5, and 2.5 processed for LH06, LH07, LH08, LH09,
LH10, and LH11 respectively. Their energy gain corrections were performed by
using {\bf xispi} with the latest makepi file {\tt
ae\_xi1\_makepi\_20110621.fits} in CALDB. All XIS data sets were observed with
the normal clocking and the $3\times 3$ or $5\times 5$ editing mode. The
Spaced--raw Charge Injection (SCI) was adopted except for the LH06 observation.
The observation LH06 was also analyzed in \citet{yoshino09}, and our result is
consistent with theirs (see the details below).

\section{Data Screening}

We start from the XIS1 cleaned event files that are processed with standard
data screening criteria. To avoid high energy particle penetration due to the
Earth's magnetic field, we extracted the data when Cut Off Rigidity (COR2) was
larger than 8 GV $c^{-1}$. Two bright sources whose flux was larger than $1.0
\times 10^{-14}$ erg cm$^{-2}$ s$^{-1}$ in the 0.5 -- 2.0 keV band were removed
by circles with diameters $3'.0$ and $4'.0$ centered at
$(\alpha=162^{\circ}.811,\ \delta=57^{\circ}.271)$ and $(163^{\circ}.005,\
57^{\circ}.178)$.

After the data were screened with the above criteria, we additionally removed
the intervals when geocoronal SWCX-induced lines and the neutral
O\emissiontype{I} K$\alpha$ line in the Earth's atmosphere were enhanced
(screening procedures were described in $\S3.1$ and $\S3.2$). Table
\ref{tab:obslog} shows the exposures of the total cleaned events and the
screened ones, and what screening was applied in each observation.

\subsection{Removal of the Geocoronal SWCX X-ray Emission}
\label{subsec:gswcx}

\begin{figure*}
\begin{tabular}{lr}
\begin{minipage}{0.5\hsize}
	\FigureFile(80mm,50mm){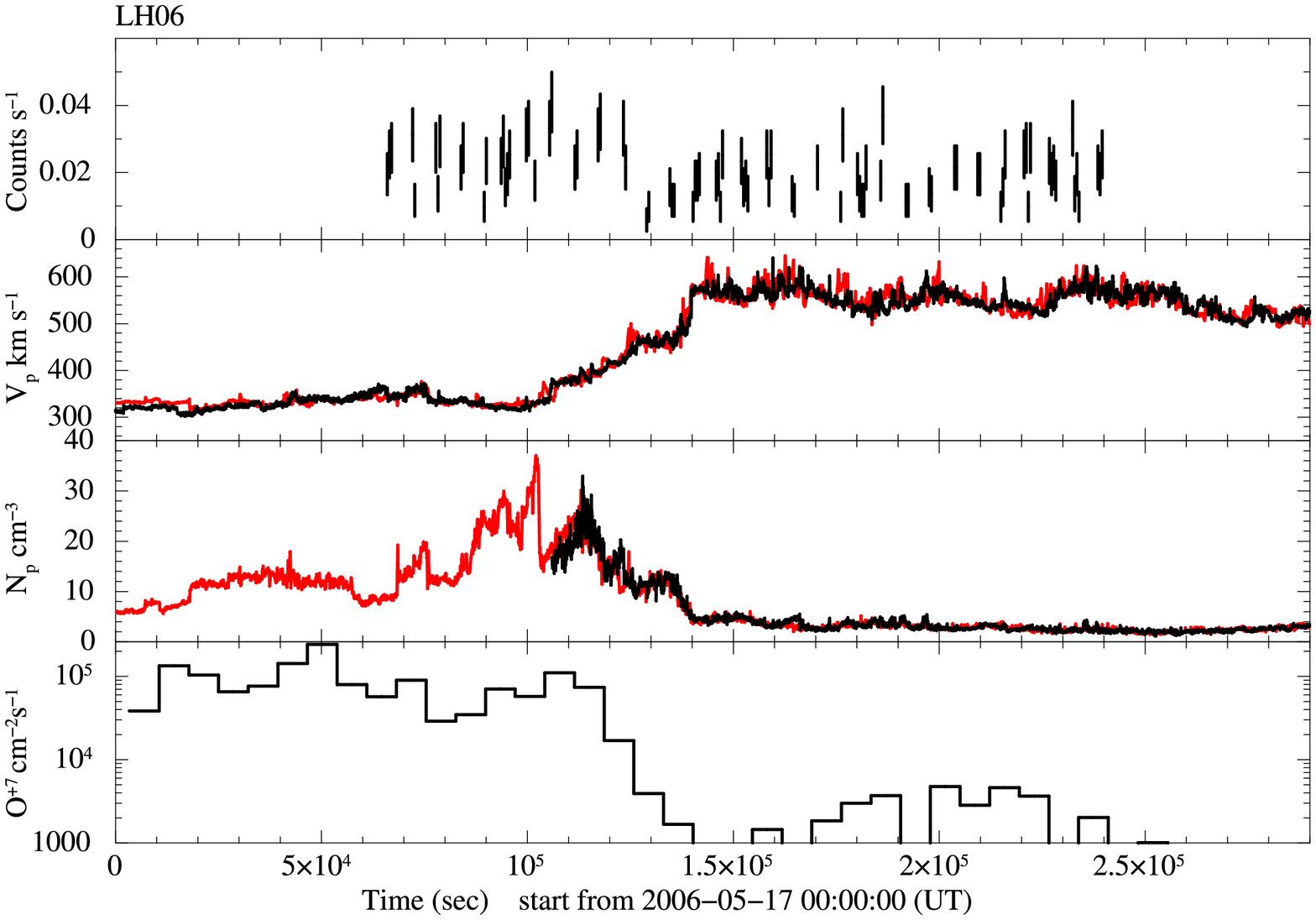}\\
\end{minipage} \ \ 
\begin{minipage}{0.5\hsize}
	\FigureFile(80mm,50mm){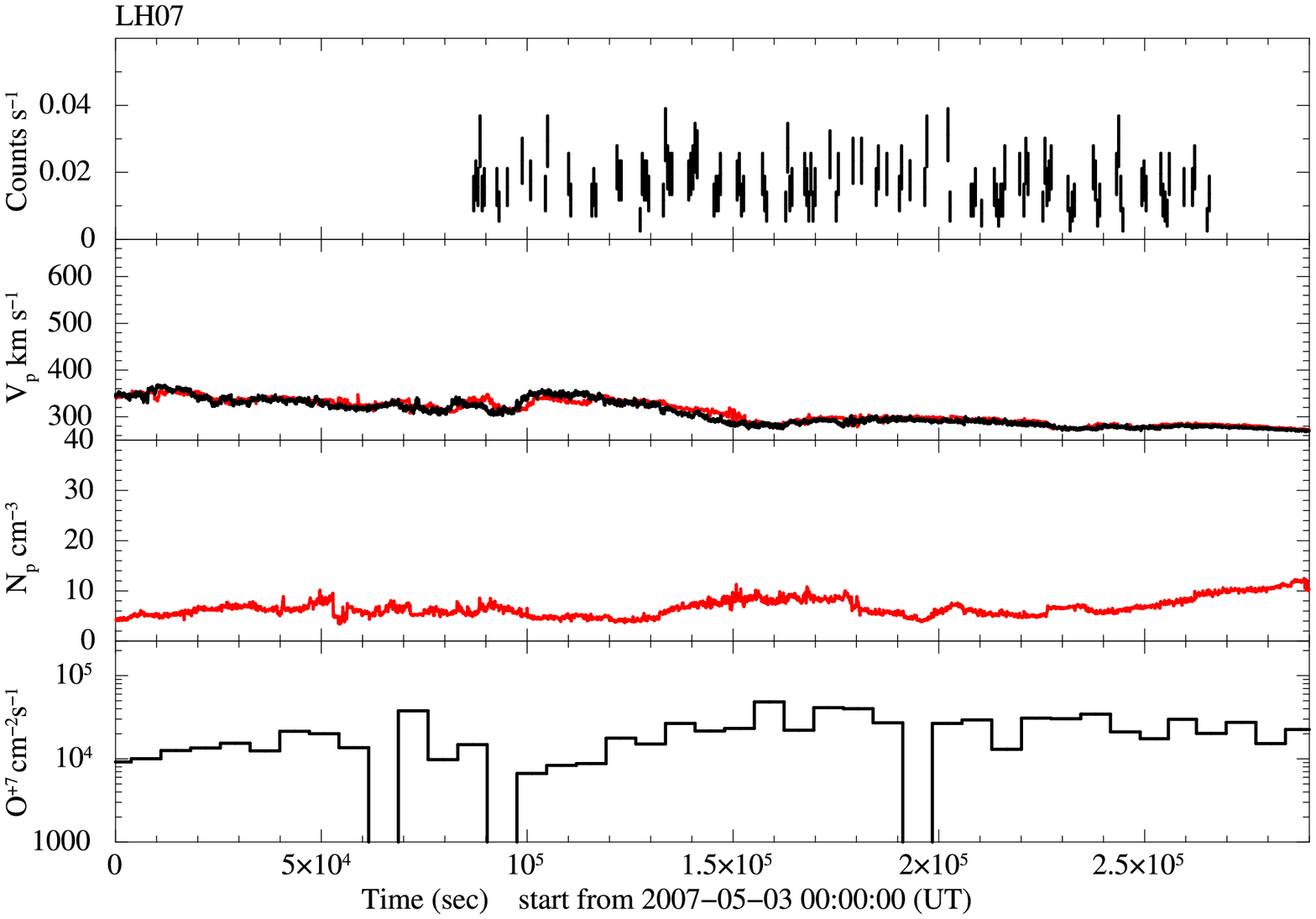}\\
\end{minipage} \\ \\ 
\begin{minipage}{0.5\hsize}
	\FigureFile(80mm,50mm){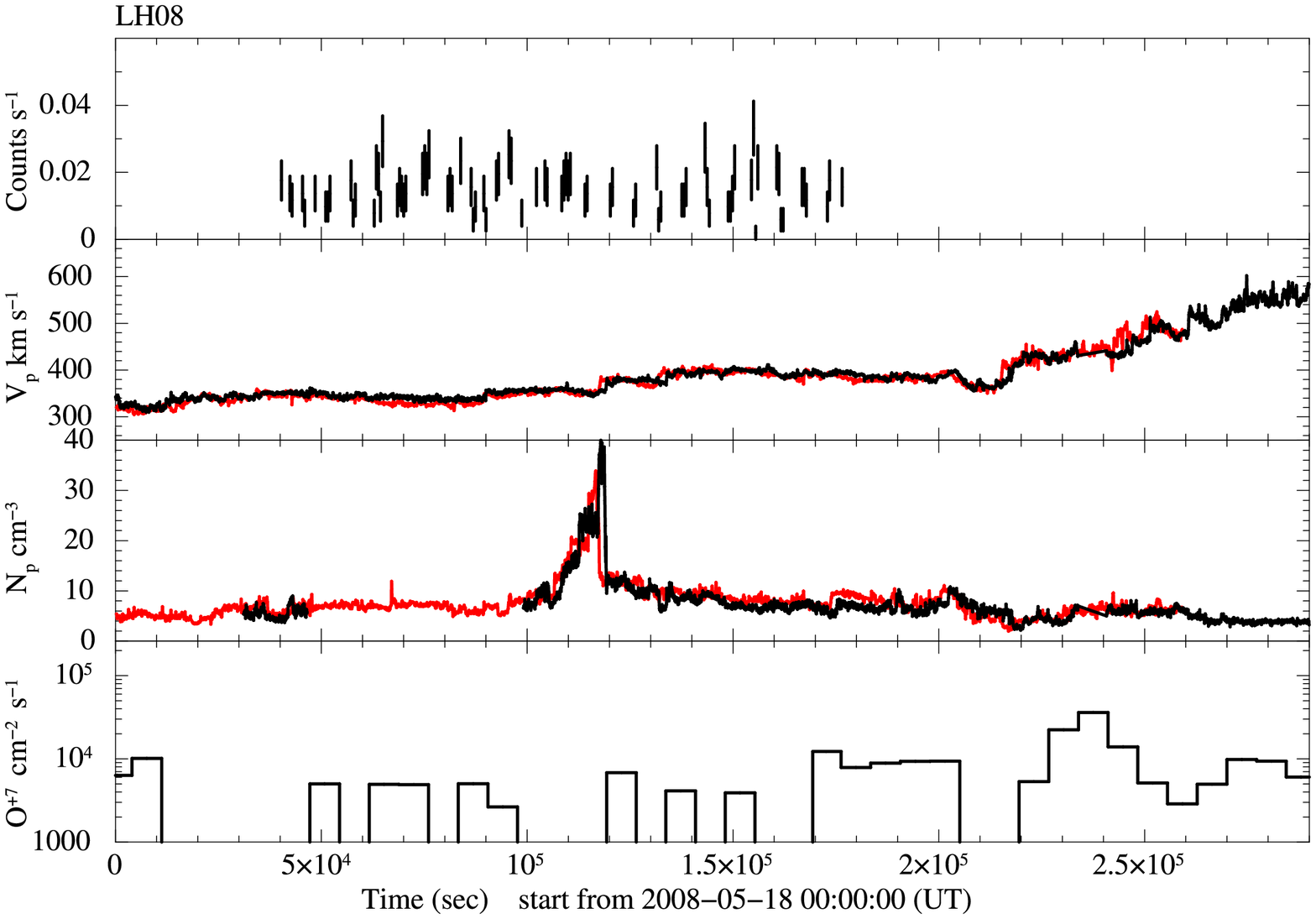}\\
\end{minipage} \ \ 
\begin{minipage}{0.5\hsize}
	\FigureFile(80mm,50mm){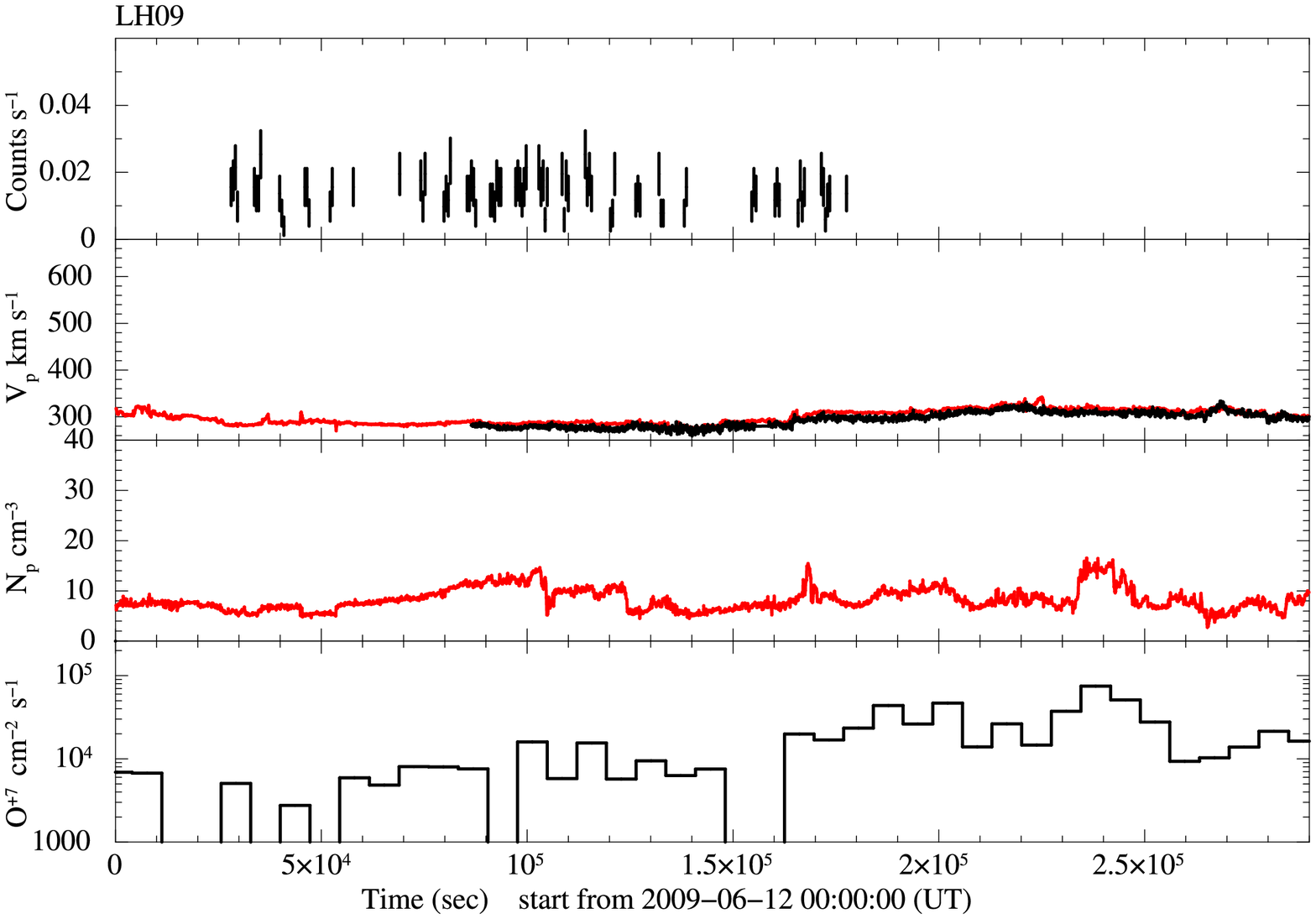}\\
\end{minipage} \\ \\ 
\begin{minipage}{0.5\hsize}
	\FigureFile(80mm,50mm){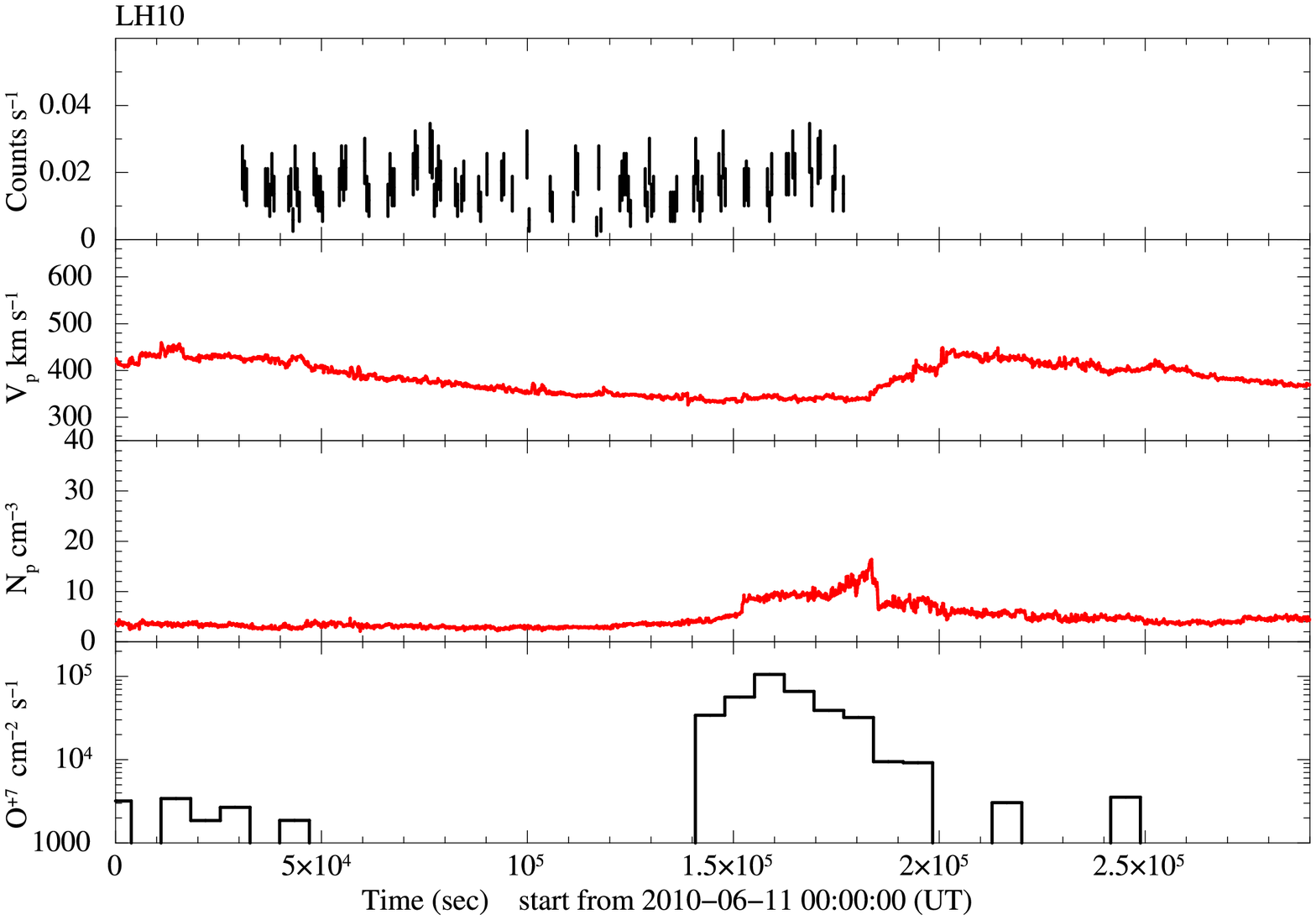}\\
\end{minipage} \ \ 
\begin{minipage}{0.5\hsize}
	\FigureFile(80mm,50mm){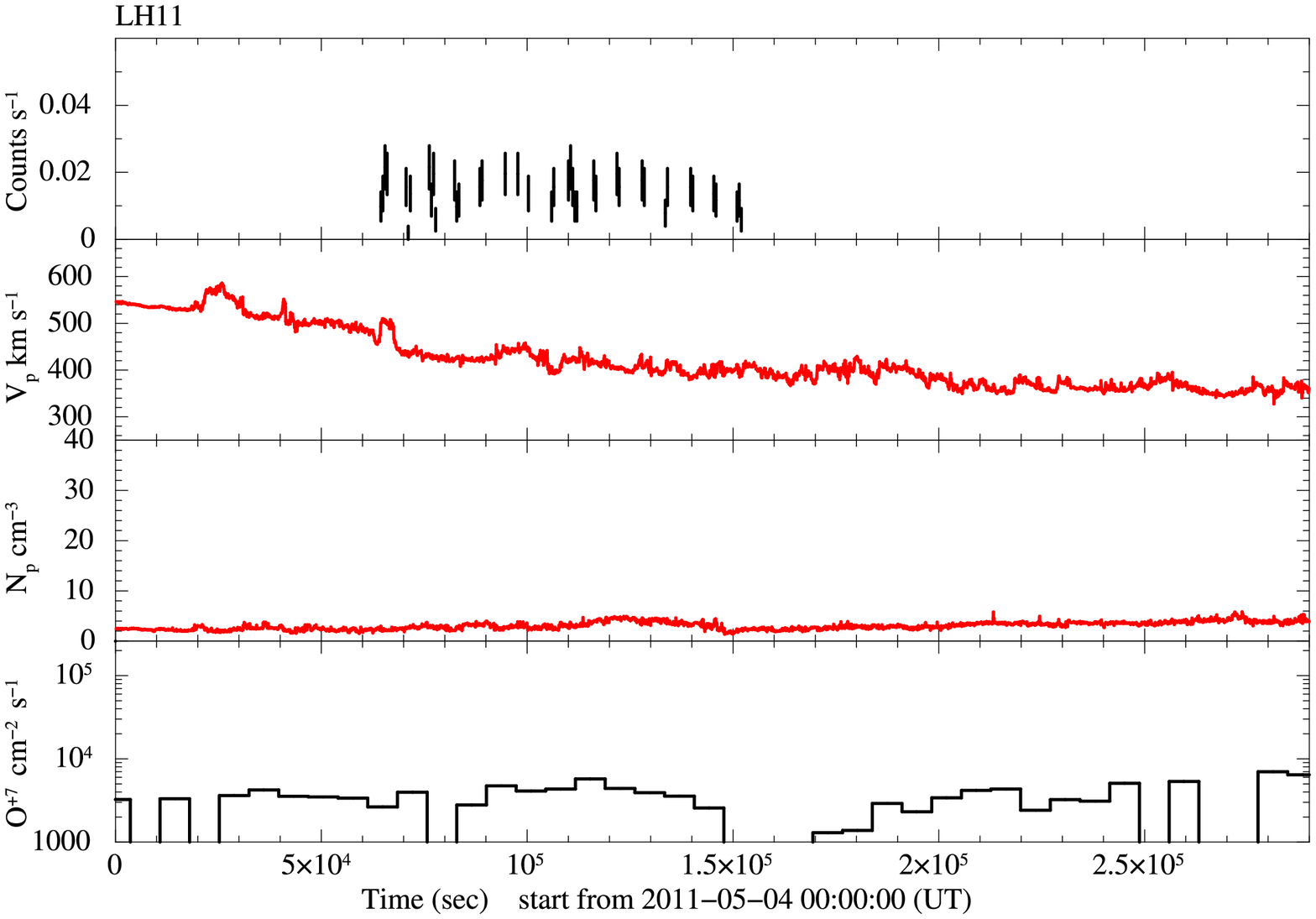}\\
\end{minipage} \\
\end{tabular} 
\caption{Correlation between soft X-ray light curves and solar wind
 parameters
during the observations; 
Suzaku/XIS1 512 s binned 0.5 -- 0.7 keV  light curve (top), 
ACE/SWEPAM (black) 64 s and WIND/SWE (red) 96 s time resolution solar
 wind proton velocity (2nd row), 
solar wind proton density (3rd row), and 
ACE/SWICS--SWIMS 2 hour averaged solar wind O$^{+7}$ ion flux (bottom).}
\label{fig:lc_proton_ion}
\end{figure*}

The time scale of geocoronal SWCX variations is about tens of ksec to a few
days (\cite{snowden04,fujimoto07}). Therefore, we can assume that when
variations reach a minimum intensity during each observation the spectrum is
the least contaminated by geocoronal SWCX. The intensity of the geocoronal
SWCX-induced line depends on the product of the neutral density in the
atmosphere and the solar wind ion flux. A typical scale hight of the geocoronal
neutrals is $\sim 8.2 R_{\rm E}$ from the Earth's surface (\cite{estrg03}), and
the penetration depth of solar wind ions to the atmosphere is determined by the
interplanetary plasma conditions (\cite{yoshino09}). Here we explain the
specifics of how to determine the data reduction criteria.
 
First, we checked the solar wind data of monitoring satellites;
ACE/SWEPAM\footnote{http://www.srl.caltech.edu/ACE/ASC/level2/\ \ \ \
lvl2DATA\_SWEPAM.html} or
WIND/SWE\footnote{http://web.mit.edu/space/www/wind/wind\_data.html} for the
solar wind protons, and
ACE/SWICS--SWIMS\footnote{http://www.srl.caltech.edu/ACE/ASC/level2/\ \ \ \
lvl2DATA\_SWICS-SWIMS.html} for the solar wind heavy ions, respectively. The
data of ACE/SWICS--SWIMS were only used when the quality flags of solar wind
parameters were equal to 0 (meaning ``Good quality''). In these observation
intervals, there were some missing ACE Level 2 data, for which we compensated
with the solar wind proton data measured by WIND.
Figure~\ref{fig:lc_proton_ion} shows the X--ray light curves in the 0.5 -- 0.7
keV band, where O\emissiontype{VII} and O\emissiontype{VIII} emission lines are
dominant, of Suzaku XIS1, the solar wind proton density and velocity with
ACE/SWEPAM (black) and WIND/SWE (red), and solar wind O$^{+7}$ ion flux with
ACE SWICS--SWIMS during each observation. Both ACE and WIND are orbiting around
L1 ($\sim235$ $R_{\rm E}$ sunward of Earth), and there is a delay in the solar
wind propagation from L1 to the Earth. We shifted the solar wind data by using
the velocity averaged over each observation. We compared energy spectra
extracted from different intervals sorted by the proton flux in each
observation, in order to judge the necessity of screening. Note that the
statistical errors in the X-ray light curves are too large to visually identify
the influence of geocoronal SWCX.

During LH07, LH09, and LH11, the proton fluxes, expressed as a product of the
density and the velocity, were always less than $4.0\times 10^{8}$
cm$^{-2}$s$^{-1}$. According to the solar wind monitoring at L1, both LH07 and
LH09 occurred during slow solar wind conditions, and LH11 was in temporary fast
conditions (typical properties of the slow/fast solar winds are summarized in
\citet{kout06}, Table 1). These were not screened by solar wind flux.

In LH06 and LH10, both proton and ion fluxes were enhanced by the passage of a
corotating interaction region (CIR) around the Earth, in which slow solar wind
is compressed by fast wind. We compared the O\emissiontype{VII} line
intensities between the CIR passage and the rest of the observation interval.
The O\emissiontype{VII} intensities estimated from spectral fitting (details
are described in the next section) were $5.6_{-1.1}^{+1.1}$ (CIR passage) /
$2.6_{-0.7}^{+0.8}$ (the rest) LU in LH06, and $7.7\pm 1.3$ (CIR passage) /
$6.1\pm1.1$ (the rest) LU in LH10, respectively. We used only the data when the
solar wind was consistently slow or fast (i.e.\ from 90 ks after the start of
LH06, and within 110 ks from the start of LH10).

Although the proton density data presented a steep increase between 60 ks and
100 ks from the beginning of LH08, there was no enhancement in both the
O$^{+7}$ flux and O\emissiontype{VII}. Estimated O\emissiontype{VII}
intensities are $2.7_{-1.3}^{+1.5}$ (60 ks to 100 ks) / $3.0_{-0.9}^{+1.0}$
(the rest) LU respectively. Therefore, we did not screen the LH08 data by the
solar wind flux.

Second, we calculated the structure of the geomagnetic field towards the Suzaku
line of sight (LOS) using the Geopack 2008 (\cite{tsyganenko05}). In all
observations, the distance of the Earth--center to the magnetopause (ETM)
varied from 1.5 to 15.0 $R_{\rm E}$. However, the effect of low ETM distance on
the energy spectrum was only confirmed during the passage of a CIR in LH06 that
had been already removed. In this analysis, we did not screen the data based on
the ETM distance.

\subsection{Removal of the Neutral O\emissiontype{I} K$\alpha$ Line.}
\label{subsec:neu_O}

We checked the mixing of neutral oxygen lines to the O\emissiontype{VII} line.
Solar X-ray photons scattered with neutral oxygen in Earth's atmosphere create
the O\emissiontype{I} K$\alpha$ line emission (centroids are $E_{\rm
O_{2}}=525$ eV and $E_{\rm O}=540$ eV). In accordance with previous studies
(\cite{smith07}, \cite{miller08}, and \cite{yoshino09}), we calculated the
distributions of the neutrals in Earth's atmosphere using the Normal MSISE--00
model 2001\footnote{http://ccmc.gsfc.nasa.gov/modelweb/atmos/nrlmsise00.html}
(\cite{hedin91}).

The count rate in the 0.5--0.6 keV band increased with the neutral column
density only in LH07. We decided to use the X-ray events of LH07 when the
neutral oxygen column density was less than $1.0 \times 10^{14}$ cm$^{-2}$ for
observation intervals on Earth's dayside, or $1.0 \times 10^{15}$ cm$^{-2}$ on
the nightside. The spectral differences in the 0.5--0.6 keV band were also seen
in LH11 when the satellite orbited between the Earth's day and night side. It
is thought to be an effect of the neutral scattering occurring in the
atmosphere very close to Suzaku. We only used the LH11 data for spectral
analysis when Suzaku was on the nightside.

In addition, a GOES C4.2 class X-ray flare was observed on May 5th 11:00 UT
during LH07, and an M2.0 class flare was observed on June 12th 9:00 UT during
LH10. However, there was no significant O\emissiontype{I} K$\alpha$ enhancement
in either the Suzaku light curve or the energy spectrum, so we included the
data acquired during the flare arrivals.

\bigskip

\section{Spectral Analysis}

\label{sec:analysis}

\subsection{Response and Background Files}

We created a Redistribution Matrix File (RMF) for conversion between the Pulse
Invariant (PI) channel spectrum and the energy spectrum, and an Auxiliary
Response File (ARF) for the spectral fitting, using the Suzaku FTOOLS software
{\bf xisrmfgen} ver.~2009--02--28 and {\bf xissimarfgen} ver.~2010--11--05,
respectively (\cite{ishisaki07}). The emission source of the ARF file is
assumed to be a uniform sky with radius of $20'$.

A Non X-ray Background (NXB) was constructed using the software {\bf xisnxbgen}
ver.~2008--03--08 (\cite{tawa08}). In order to confirm the reproducibility of
{\bf xisnxbgen}, we compared count rates of the calculated NXB with those in
each Lockman Hole observation above 12~keV, where the XRT does not reflect
X-ray photons. The count rates were consistent within 10\%. These results were
consistent with other papers (\cite{masui09} and \cite{yoshino09}). To correct
the slight difference of the calculated NXB and the observed background, we
scaled the normalization of NXB spectrum to equalize its count rate above 12
keV to that of each LH observation.

\subsection{Emission Model Description}
\label{sec:analysis_model}

The purpose of this spectral analysis is to obtain the O\emissiontype{VII}
K$\alpha$ (0.57 keV) and O\emissiontype{VIII} Ly$\alpha$ (0.65 keV) line
intensities precisely, and scrutinize the variability of them among different
observations. However, there are some contributions at these line energies from
other diffuse emission components that cannot be resolved with the CCD
resolution. Therefore, we need to estimate this contribution by using a certain
range of energy spectra.

The diffuse emission model consists of three components; the extragalactic
unresolved point sources (cosmic X-ray background; CXB), the Galactic halo gas,
and a blend of the ``local hot bubble'' (LHB) and SWCX. The emission of CXB,
which is the dominant component in 2--5~keV, is represented by double broken
power laws. Their photon indices are ${\mit \Gamma} = 1.96$ and ${\mit \Gamma}
= 1.54$ below the folding energy at 1.2 keV and ${\mit \Gamma} = 1.40$ above
it. Following \citet{smith07}, we normalized the broken power law component
with photon index of ${\mit \Gamma} = 1.54$ to 5.7 photons s$^{-1}$ keV$^{-1}$
sr$^{-1}$ at 1 keV. Galactic halo gas, the hot interstellar medium in the
Galaxy, is represented by a thin thermal plasma with $kT \sim 0.25$ keV,
absorbed by the galactic neutrals (\cite{kuntz00}). The blend of LHB ($\sim
10^{6}$ K gas surrounding our Solar system) and SWCX is represented by an
unabsorbed thin thermal plasma with $kT \sim 0.1$ keV. Though the emission
process and fine structures of SWCX induced lines are quite different from a
collisional ionization equilibrium (CIE) plasma (\cite{snowden09}), we treated
them as a single CIE plasma in this analysis in accordance with many previous
works (\cite{smith07}, \cite{yoshino09}, and \cite{henly10}). For the thin
thermal plasma emission, the {\it APEC\,} model (with AtomDB ver.2.0.1) is
applied to estimate both the Galactic halo gas and the LHB$+$SWCX emissions.

\subsection{Spectral Fitting in 0.4--5.5 keV Band}
\label{sec:analysis_fit_wide}

\begin{figure*}
\begin{center}
	\FigureFile(160mm,100mm){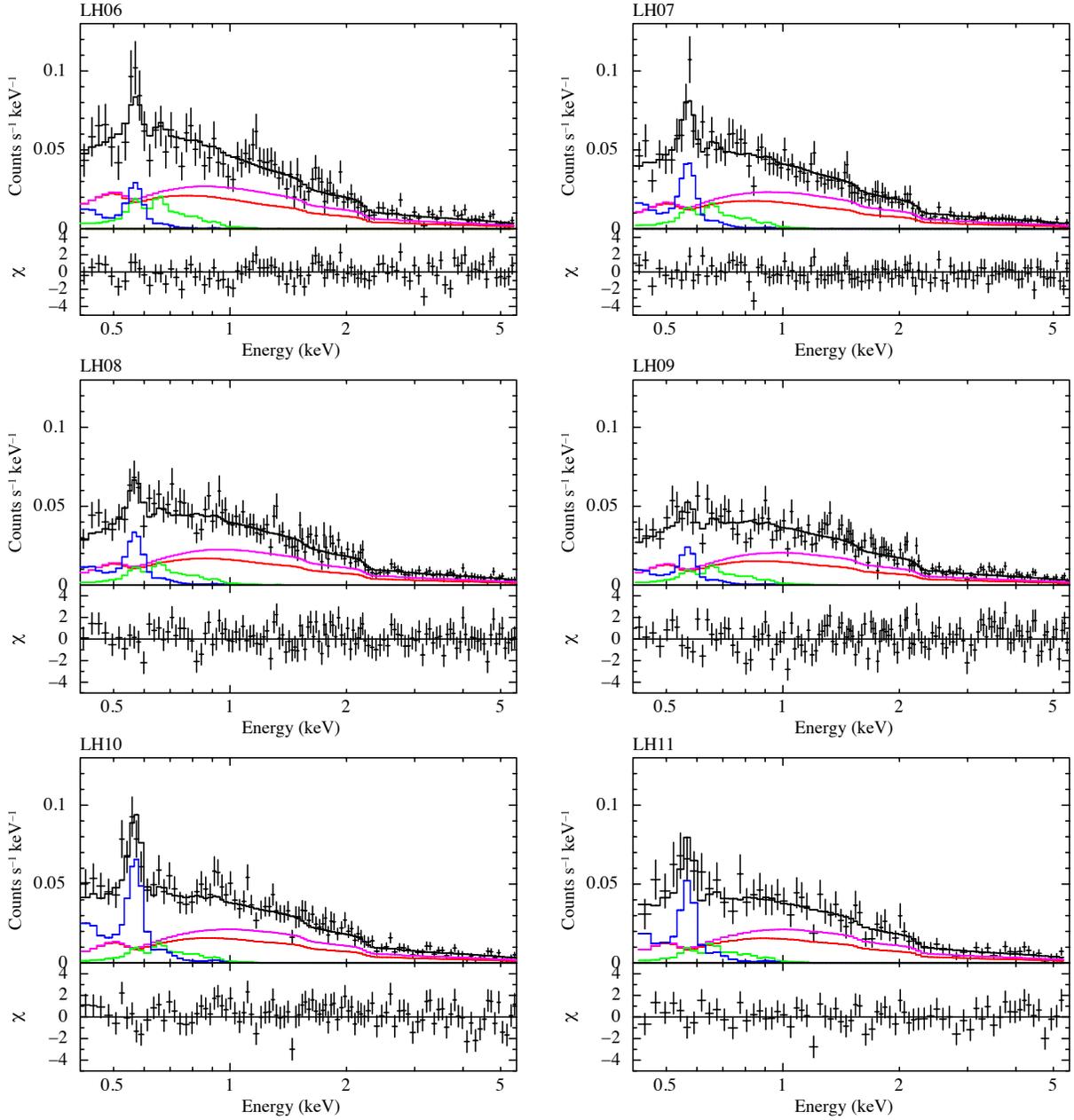}
\end{center}
\caption{0.4 -- 5.5 keV Suzaku XIS1 (Back Illuminated CCD) spectra and best fit emission models of the blank field toward the
Lockman hole from 2006 to 2011 convolved with the CCD and telescope responses (top panel) and residual of the fit (bottom panel). 
Black crosses show the observed spectra. Step lines show the best fit  models; total (black),  Galactic halo (green), 
LHB$+$SWCX (blue), CXB with ${\mit \Gamma} = 1.54$ (magenta), 
and CXB with ${\mit \Gamma} = 1.96$ (red) respectively.}
\label{fig:spec_tae}
\end{figure*}

To estimate the oxygen line intensities, we first tried to fit the spectra
using a broad energy range from 0.4 to 5.5 keV. Figure \ref{fig:spec_tae} shows
the Lockman Hole spectra with their best fit double broken power laws $+$
Galactic halo $+$ (LHB $+$ SWCX) models. All six spectra are fitted
simultaneously. First, we tried to fit the spectra with the same normalization
of CXB for all observations, because their pointing directions were almost the
same. The result showed the 90 \% confidence ranges of the temperature and the
normalized halo all overlapped. Therefore, we decided to take the same value of
the halo parameters for all observations. The temperature of the LHB$+$SWCX
blend was fixed at the typical temperature $kT=0.099$ keV ($\sim
1.15\times10^{6}$ K), derived from observations of the local blank field
region, where the contribution of the halo was almost negligible due to the
large interstellar absorption (\cite{yoshino09}, \cite{hagihara10}). The best
fit values are summarized in the upper 6 rows in Table \ref{tab:spec_tae_dpow}.
The reduced $\chi^{2}$ value is 0.999 for 636 d.o.f., and all spectra were well
reproduced by this model.

\subsection{Derivation of O\emissiontype{VII} and  O\emissiontype{VIII} Line Intensities}
\label{sec:analysis_fit_ox}

After determining the model parameters of wide band spectra in
$\S$\ref{sec:analysis_fit_wide}, we tried to fit the same spectra again by
fixing the halo temperature at the best fit values, and setting zero oxygen
abundances for both the halo and LHB$+$SWCX, but inserting three Gaussian lines
at the O\emissiontype{VII} K$\alpha$, O\emissiontype{VIII} Ly$\alpha$, and
O\emissiontype{VII} K$\beta$ energies. As reported in \citet{yoshino09}, the
O\emissiontype{VIII} Ly$\alpha$ line intensity is weak and the
O\emissiontype{VII} K$\beta$ line at 666 eV cannot be resolved from
O\emissiontype{VIII} Ly$\alpha$. For this reason, it is hard to determine the
error of the O\emissiontype{VIII} Ly$\alpha$ line energy centroid in the
Lockman Hole observations. We fixed the centroid of O\emissiontype{VII}
K$\alpha$ at 567 eV, averaging the centroids of resonance (574 eV) and
forbidden (561 eV) lines, and the centroids of O\emissiontype{VIII} Ly$\alpha$
and O\emissiontype{VII} K$\beta$ at 653 eV and 666 eV respectively. The
fraction of the O\emissiontype{VII} K$\beta$ to O\emissiontype{VII} K$\alpha$
was also fixed at $8.3\, \%$ (\cite{kharchenko03}, \cite{yoshino09}). The lower
6 rows in Table \ref{tab:spec_tae_dpow} show the best fit results with this new
model, and here we define the intensity of O\emissiontype{VII} and
O\emissiontype{VIII} lines by normalization of the two inserted Gaussian lines.
According to this fitting result, the O\emissiontype{VII} intensities of the
earlier four observations were almost constant within the 90 \% statistical
error range. However, those of LH10 and LH11 are $2-3$ LU ($= {\rm photons\
s}^{-1}{\rm cm}^{-2}{\rm sr}^{-1}$) brighter than the others. The statistical
significances of the O\emissiontype{VII} intensity variation of LH10 and LH11
with respect to an average of LH06 to LH09 ($=2.99\pm0.38$ LU) are $4.5\sigma$
(LH10) and $2.3\sigma$ (LH11), respectively.

When we use an APEC model with zero O abundance, it is implicitly assumed that
the continuum by O atoms is negligible. To confirm that the contribution of the
continuum is small, we also fitted the spectra with a thermal plasma model
including the continuum by O atoms but no O\emissiontype{VII} K$\alpha$,
O\emissiontype{VIII} Ly$\alpha$ and O\emissiontype{VII} K$\beta$ lines, along
with three Gaussian lines representing the O emission lines. The normalization
of the three Gaussian lines did not change by more than a few \%. Hence, we
conclude that the contribution of the continuum is small. Note that continuum
emission is expected if the origin is a thermal plasma, while no continuum is
expected if the origin is SWCX. Our results suggest that the intensity of the
three lines can be precisely determined regardless of the continuum assumption,
i.e., the origin of the emission.

\subsection{Systematic Uncertainties of O\emissiontype{VII} / O\emissiontype{VIII} Lines }
\label{subsec:sys_uncertainty}

The absolute intensities of O\emissiontype{VII} / O\emissiontype{VIII} lines
strongly depend on the X-ray detection efficiency and emission model
(\cite{yoshino09}). First, we checked the uncertainty due to the contaminant
thickness on the XIS optical blocking filter. The systematic uncertainty of
contaminant thickness is estimated to be $\pm 10 \%$ of the nominal value in
the CALDB files (Suzaku technical
description\footnote{http://www.astro.isas.jaxa.jp/suzaku/doc/suzaku\_td/} ).
We changed the efficiency described in the ARF using the software {\bf
xiscontamicalc} version 2010--11--05 according to the thickness uncertainty.
These systematic errors are smaller than the 90 \% statistical ones in Table
\ref{tab:spec_tae_dpow}, and we only show the results with the ARF of nominal
contamination thickness. If the O\emissiontype{VII} lines of LH10 and LH11 are
equal to the early four observations, the thickness must be overestimated by
$50\%$ and $40\%$ of the nominal values in May 2010 and June 2011, respectively
(according to the latest CALDB {\tt ae\_xi[1-3]\_contami\_20091201.fits}).

Second, we checked the contribution of the uncertainty of the CXB spectrum
below 2 keV to the determination of oxygen line intensities. We tried to use a
single power law model with photon index ${\mit \Gamma}=1.4$ for the CXB
emission instead of double broken power laws. The oxygen lines systematically
became $0.3-0.4$ LU larger than in the case of using the double broken power
laws CXB model. However, the differences between them were not dependent on the
CXB model.

\begin{table*}[htbp]
\begin{center}
\caption{Results of spectral fitting with double broken power laws CXB $+$ Galactic halo$+$ (LHB$+$SWCX) models.}
{\tabcolsep = 1.0mm
\begin{tabular}{cccccccc}
\hline
\hline
{\small component} & {\small CXB} &  \multicolumn{2}{c}{\small Galactic Halo}
& {\small LHB$+$SWCX} & {\small O\emissiontype{VII}} & {\small O\emissiontype{VIII}} & \multirow{4}{*}{$\chi^{2}/{\rm d.o.f.}$} \\ 
{\small \it model} & {\small \it  phabs$^{\ast}$(bknpwls$^{\dagger}$)}
& \multicolumn{2}{c}{\small \it  phabs$^{\ast}$(APEC)} 
& {\small \it APEC$^{\,\ddagger}$} & {\small \it  gaussian$^{\S}$} & {\small \it  gaussian$^{\S}$} \\ 
{\small parameter} & {\small norm.} & $kT$ & {\small norm.} 
& {\small norm.} & {\small norm.} & {\small norm.} \\ 
{\small unit} & {\small $^{\parallel}$} & {\small keV} &  {\small $^{\sharp}$} 
& {\small $^{\sharp}$} & {\small $^{\ast\ast}$} & {\small $^{\ast\ast}$}  \\ \hline 
{\small LH06} & \multirow{6}{*}{\small $4.1\pm0.2$} 
& \multirow{6}{*}{\small $0.216_{-0.029}^{+0.025}$} &
\multirow{6}{*}{\small $1.7_{-0.4}^{+1.1}$} &
{\small $9.1_{-5.3}^{+4.0}$} & $-$ & $-$ & 
\multirow{6}{*}{\small $635.2/636$} \\
{\small LH07} & 
& {\small $_{}^{}$} &
{\small $_{}^{}$} &
{\small $18.4_{-5.4}^{+4.1}$} & $-$ & $-$ & 
\\
{\small LH08} & 
& {\small $_{}^{}$} &
{\small $_{}^{}$} &
{\small $17.4_{-5.5}^{+4.3}$} & $-$ & $-$ & 
\\
{\small LH09} & 
& {\small $_{}^{}$} &
{\small $_{}^{}$} &
{\small $14.7_{-5.6}^{+4.5}$} & $-$ & $-$ & 
\\
{\small LH10} & 
& {\small $_{}^{}$} &
{\small $_{}^{}$} &
{\small $39.7_{-6.5}^{+5.7}$} & $-$ & $-$ & 
\\
{\small LH11} & 
& {\small $_{}^{}$} &
{\small $_{}^{}$} &
{\small $31.8_{-8.9}^{+8.4}$} & $-$ & $-$ & 
\\
\hline
{\small LH06} & \multirow{6}{*}{\small $4.1\pm0.2$} 
&  \multirow{6}{*}{\small $0.216$ (fixed)} &
 \multirow{6}{*}{\small $2.1\pm 0.8$} &
{\small $9.0\pm 8.4$} &
{\small $2.55 \pm 0.74$} &
{\small $0.21\ (< 0.62)$} &
\multirow{6}{*}{\small $615.9/625$} \\
{\small LH07} & 
& {\small $ $ } &
{\small $_{}^{}$} &
{\small $20.7\pm 9.7$} &
{\small $3.68\pm 0.72$} &
{\small $0.61\pm 0.37$} 
\\
{\small LH08} & 
& {\small $ $ } &
{\small $_{}^{}$} &
{\small $27.5\pm 10.3$} &
{\small $3.03\pm 0.77$} &
{\small $0.94\pm 0.41$} 
\\
{\small LH09} & 
& {\small $ $} &
{\small $_{}^{}$} &
{\small $25.0\pm 10.9$} &
{\small $2.69\pm 0.80$} &
{\small $0.77\pm 0.42$} 
\\
{\small LH10} & 
& {\small $ $ } &
{\small $_{}^{}$} &
{\small $60.1\pm 13.7$} &
{\small $6.06\pm 1.07$} &
{\small $0.86\pm 0.49$} 
\\
{\small LH11} & 
& {\small $ $} &
{\small $_{}^{}$} &
{\small $42.0\pm 22.5$} &
{\small $5.28\pm 1.60$} &
{\small $0.97\pm 0.76$} 
\\
\hline
\multicolumn{8}{l}{\small $^\ast$ 
 Absorption column density is fixed at $N_{\mathrm {H}} = 5.8\times 10^{19} {\mathrm {cm}}^{-2}$ (\cite{kalb05}).} \\ 
\multicolumn{8}{l}{\small $^\dagger$ 
 Normalization of the power law with  ${\mit \Gamma}=1.54$ 
 is also fixed at $ 5.7\ {\rm photons~s}^{-1}{\rm cm}^{-2}{\rm keV}^{-1}{\rm sr}^{-1}$@1keV.  } \\
\multicolumn{8}{l}{\small $^\ddagger$ 
See details in the text in \S\ref{sec:analysis_fit_wide}.} \\
\multicolumn{8}{l}{\small $^\S$ 
See details in the text in \S\ref{sec:analysis_fit_ox}.} \\
\multicolumn{8}{l}{\small $^\parallel$
The unit of  the normalization of a power law component is 
${\rm photons~s}^{-1}{\rm cm}^{-2}{\rm keV}^{-1}{\rm sr}^{-1}$@1keV. } \\
\multicolumn{8}{l}{\small $^\sharp$ 
 The emission measure integrated over the line of sight, 
 $(1/4\pi)\int n_{\mathrm e} n_{\mathrm H} ds$ in the unit of $10^{14}~{\rm cm}^{-5}~{\rm sr}^{-1}$.}\\
\multicolumn{8}{l}{\small $^{\ast\ast}$ 
The normalization of gaussian component shows the surface brightness whose unit is defined as  } \\
\multicolumn{8}{l}{\small \quad \ ${\rm L.U.}$ (Line Unit)
 $= {\rm photons\ s}^{-1}{\rm cm}^{-2}{\rm sr}^{-1}$. } 
\end{tabular}
}
\label{tab:spec_tae_dpow}
\end{center}
\end{table*}

\section{Discussion}

From the results of spectral fitting, we found that O\emissiontype{VII}
intensities toward the Lockman Hole were enhanced by $2-3$ LU in 2010 and 2011,
compared to those from 2006 to 2009, during which the intensity was constant
within 90\% statistical errors. A statistically significant variation of the
O\emissiontype{VIII} line was not detected in our analysis.

As we described in \S\ref{sec:analysis_model}, O\emissiontype{VII} emission is
considered to arise from three different origins: hot Galactic halo, LHB, and
geocoronal/heliospheric SWCX. The emission from the halo and LHB is not time
variable on the present observation time scales, and the small difference of
the field of view due to azimuthal rotations (20 \%) is not likely to produce
the observed variation.

\begin{figure}[htbp]
\FigureFile(160mm,100mm){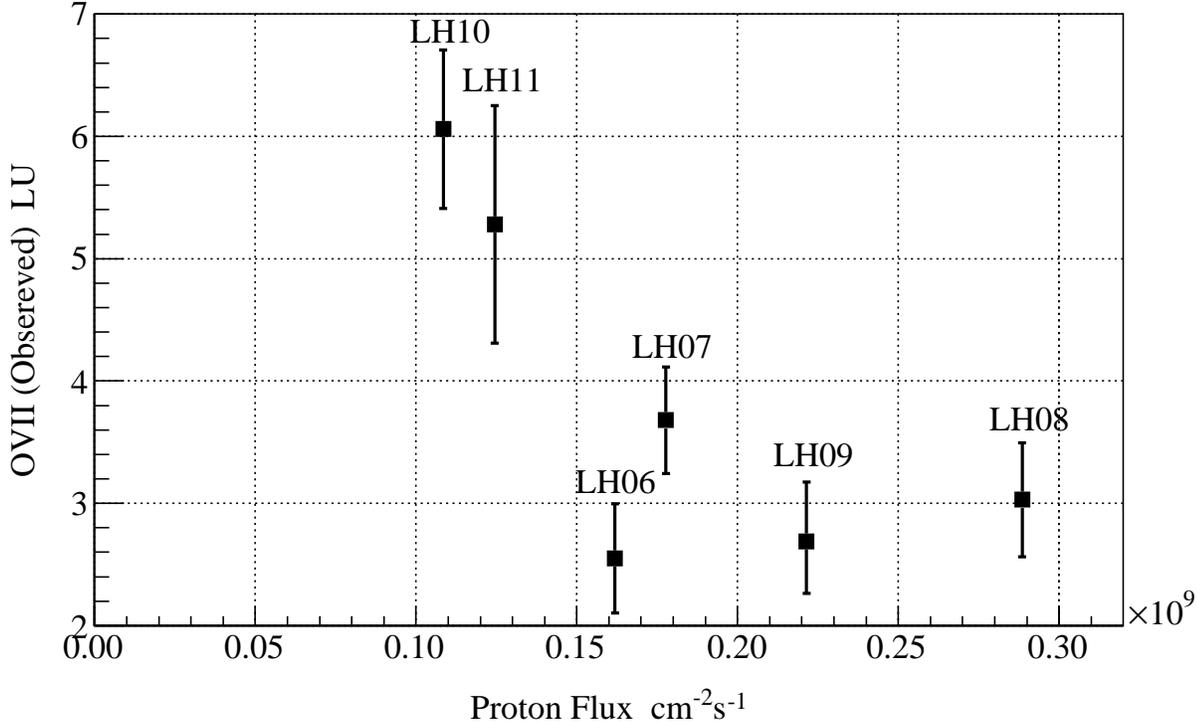}
\caption{Comparison of the  O\emissiontype{VII} line intensities obtained from the spectral fitting in 
Table \ref{tab:spec_tae_dpow} and solar wind proton flux value. Proton flux was 
represented by the average of WIND/SWE data during each Suzaku observation.}
\label{fig:OVII_vs_pflx}
\end{figure}

The data were screened for enhancements on times of a few tens of ks as much as
possible. After this correction, we may consider that most of the geocoronal
SWCX is removed. However, enhancements of the heliospheric component due to
solar wind ion flux increases in the near-Earth environment ($\sim0.5$ AU
scale) are not removed. Thus we next checked the correlation between the
O\emissiontype{VII} intensity and the solar wind flux averaged over the
observation period. The result is shown in Figure \ref{fig:OVII_vs_pflx}.
During the four Suzaku observations excluding the first and last ones (LH06 and
11), slow solar wind passed around the Earth. The ionization states of ions can
be considered to be similar for the four observations (LH07 to LH10). Therefore
we can consider that the O$^{+7}$ flux is proportional to the proton flux at
least for the four observations. Since the typical elapsed time of the
observations is two days, and the typical wind speed is $\sim 400$ km/s, the
lack of positive correlation in Figure~\ref{fig:OVII_vs_pflx} suggests that the
O\emissiontype{VII} variation is not correlated with the average ion flux on
$\sim 0.5$ AU scales.

As we discussed in $\S1$, the parallax can produce intensity variations.
However, all the observations were made between May 3 and June 13, and we find
no correlation between the O\emissiontype{VII} intensity and the day of year of
the observation. Thus the variation is not due to parallax.

We thus consider that the variation is related to variations of the ion flux or
the density of neutrals averaged along the line of sight for lengths longer
than $\sim$ 0.5 AU. A possible explanation for this is the long term variation
of the solar wind properties associated with the solar activity. Based on the
number of sunspots, the solar minimum of the Cycle 23 was around December 2008,
and then Solar Cycle 24 began. As we described in $\S1$, one of the
observational differences between the solar maximum and the minimum is the
distributions of the slow and fast solar wind in the solar corona
(\cite{mccomas08}). The direction of the LH is in the relatively high ecliptic
latitude area, at $\beta = 45.2^{\circ}$, hence the variability of the boundary
between the slow and fast winds would affect the O\emissiontype{VII} intensity.

\begin{figure}[htbp]
\begin{center}
\FigureFile(160mm,100mm){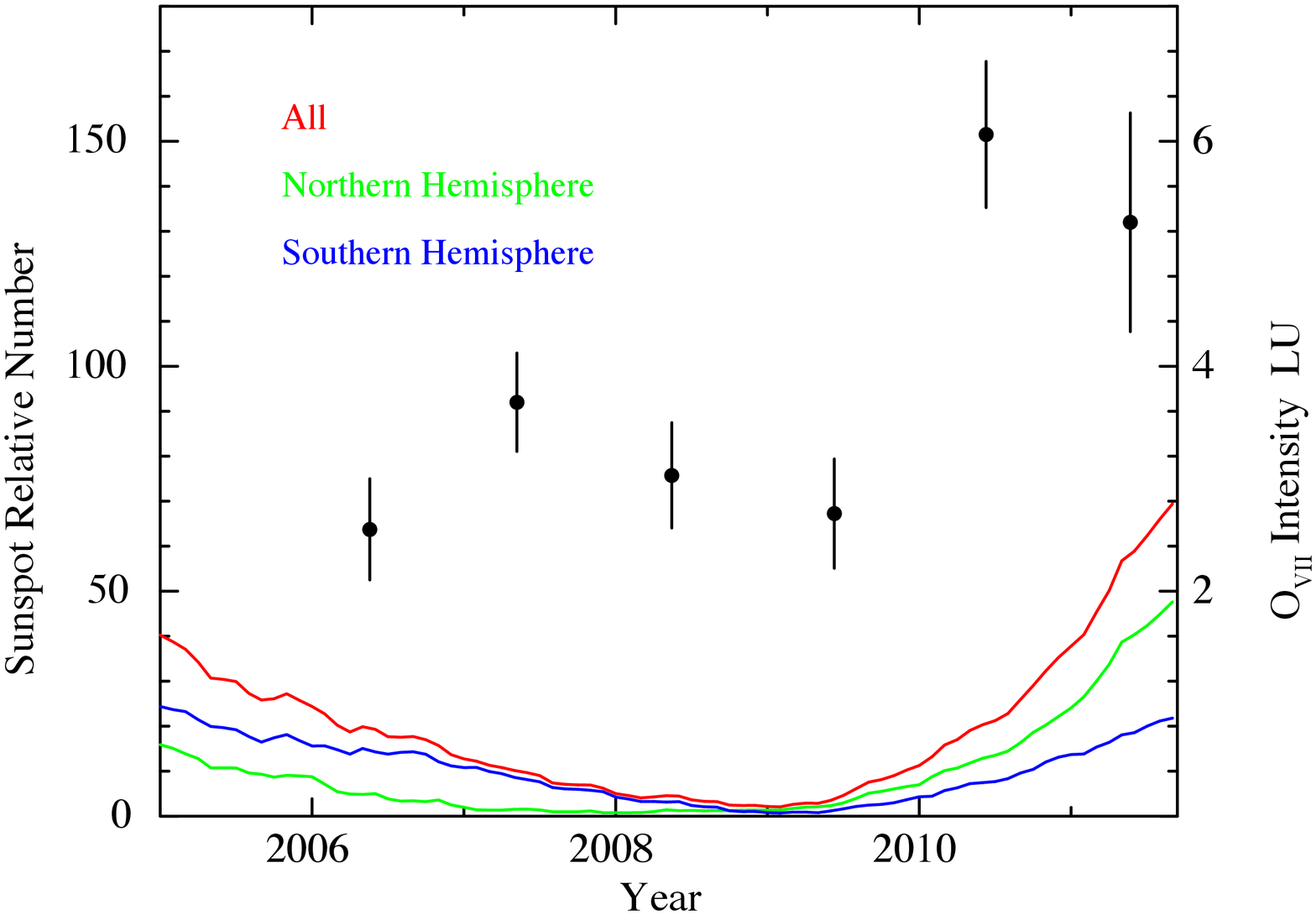}
\end{center}
\caption{Time dependences of the NAOJ relative sunspot numbers 
(red, green, and blue, left $y$ axis) and Suzaku O\emissiontype{vii} line intensities (black, right one).}
\label{fig:sunspot_ovii}
\end{figure}

Figure \ref{fig:sunspot_ovii} shows the time dependences of relative sunspot
number (NAOJ, private communication with Prof. S. Tsuneta) representing the
long term ($\sim11$-year) solar cycle together with the present
O\emissiontype{VII} intensities. Solar activities in the northern and southern
hemispheres show different time dependences. The sunspot numbers in the
southern hemisphere change symmetrically about the solar minimum, and those in
2006 are almost the same as in 2010 to 2011. On the other hand, the sunspot
numbers in northern hemisphere are less than 10 from 2006 to 2010, while they
rapidly increase after the minimum and reach 40 in June 2011. The Suzaku line
of sight (LOS) points toward the northern hemisphere of the Sun, and here we
focus on the solar activity in the northern hemisphere. The figure suggests
that the enhancement of O\emissiontype{VII} emission in 2010 is related to the
rapid increase of sunspot number in the northern hemisphere. This unusual
situation may be correlated with the north-south assymetry in Sun's polar
magnetic field reported by Hinode ({\cite{shiota12}}).

The solar wind distribution maps are obtained by the interplanetary
scintillation method (STE
lab.\footnote{http://stsw1.stelab.nagoya-u.ac.jp/ips\_data.html}). Though the
fast wind whose velocity is $>\sim600$ km s$^{-1}$ extends at high latitudes
from 2006 to 2009, the slow solar wind ($<\sim400$ km s$^{-1}$) actually
spreads throughout the northern hemisphere in 2011. If the slow/fast solar wind
boundary is at $|\beta| = 20^{\circ}$ at solar minimum, the line of sight (LOS)
of the present observations crosses the boundary and enters into the fast-wind
region at $\sim0.5$ AU from the Earth. Beyond that point, the LOS stays in the
fast wind region. On the other hand, if we assume the boundary is at $|\beta| =
50^{\circ}$ at the solar maximum, the LOS is in the slow wind all the way to
the outer boundary of the Heliosphere.

Next, we estimated the SWCX O\emissiontype{VII} intensity following the model
developed by \citet{kout06}. In their model, both solar wind ions and
interstellar neutrals distribute self-consistently by considering charge
exchange loss processes; the density of the solar wind ions basically has an
$r^{-2}$ dependence from the Sun, and the densities of interstellar hydrogen
and helium were simulated from the so-called classical ``hot model''
(\cite{lallement85}). We integrated the heliospheric SWCX emission along the
LOS within 90 AU from the Earth. All of the parameters used in this model are
same as the \citet{kout06}: solar wind proton density is 6.5/3.2 cm$^{-3}$,
velocity 400/750 km s$^{-1}$, O/H abundance ratio 1/1780 / 1/1550, and O$^{+7}$
ionization fraction 0.20/0.03 for the slow/fast wind at 1 AU respectively. In
this case, the model predicts O\emissiontype{VII} line $2.5/1.7$ LU assuming
the solar maximum/minimum conditions for the neutral distribution (Note that
the increase of O\emissiontype{VII} intensity is smaller than that expected
when we just change the beta angle of fast/slow wind boundary, because the
density of neutrals decreases due to strong photoionization by the solar UV
photons and charge exchange by the solar wind protons.) Thus this model can
explain about one third of the observed O\emissiontype{VII} intensity
variations. There are uncertainties in the solar wind parameters, in
particular, of high ecliptic latitude directions. The factor of about three
discrepancy may be explained by those uncertainties.

Because the solar wind contains ions other than O$^{+7}$, the SWCX induced
X-ray emission should also contain emission other than O\emissiontype{VII}.
However, those lines were too weak. O\emissiontype{VIII} emission was
positively detected, although its temporal variation was not significantly
detected. From the \citet{kout06} maps we estimate that the OVIII line intensity
during solar maximum is $0.8$ LU and during solar minimum $0.4$ LU. The upper
limit of the measured OVIII intensity variation ($0.75$ LU) is consistent with
the model predictions.

\bigskip


We are grateful to Dr.\@ Dimitra Koutroumpa for showing us the details of her
model and a number of fruitful suggestions as the referee. We also thank to the
Suzaku, ACE, and WIND team for providing the data. This work is partly
supported by a Grants-in-Aid for Scientific Research from JSPS (Project Number
: 09J08405, 21340046 and 22111513).


\end{document}